\documentclass[pra,twocolumn,floatfix,notitlepage,superscriptaddress,longbibliography]{revtex4-1}
\usepackage{amsmath}
\usepackage{amsfonts}
\usepackage{bbm}
\usepackage{bm}
\usepackage{graphicx}
\usepackage[bookmarks,bookmarksopen,bookmarksdepth=2]{hyperref}
\hypersetup{
    colorlinks=true,
    citecolor=green,
    linkcolor=green,
    filecolor=magenta,
    urlcolor=blue,
}
\usepackage{mathtools}
\usepackage{mathrsfs}
\usepackage{physics}
\usepackage{dsfont}
\usepackage{etoolbox}
\usepackage{datetime2}
\usepackage{quantikz}
\usepackage{ulem}

\usepackage[utf8]{inputenc}

\newcommand{\nocontentsline}[3]{}
\newcommand{\tocless}[2]{\bgroup\let\addcontentsline=\nocontentsline#1{#2}\egroup}
\makeatletter
\DeclareRobustCommand{\element}[1]{\@element#1\@nil}
\def\@element#1#2\@nil{%
  #1%
  \if\relax#2\relax\else\MakeLowercase{#2}\fi}
\makeatother

\newcommand*{\TPT}{{}^{3}\mathrm{P}_{2}}
\newcommand*{\SSZ}{{}^{1}\mathrm{S}_{0}}
\begin{document}
\widetext

\title{Qudit entanglers using quantum optimal control}

\author{Sivaprasad Omanakuttan}
\email[]{somanakuttan@unm.edu}
\author{Anupam Mitra}
\affiliation{Center for Quantum Information and Control (CQuIC), Department of Physics and Astronomy, University of New Mexico, Albuquerque, New Mexico 87131, USA}
\author{Eric J. Meier}
\affiliation{Materials Physics and Applications Division, Los Alamos National Laboratory, Los Alamos, New Mexico 87545}
\author{Michael J. Martin }
\affiliation{Materials Physics and Applications Division, Los Alamos National Laboratory, Los Alamos, New Mexico 87545}
\affiliation{Center for Quantum Information and Control (CQuIC), Department of Physics and Astronomy, University of New Mexico, Albuquerque, New Mexico 87131, USA}
\author{Ivan H Deutsch}
\email[]{ideutsch@unm.edu}
\affiliation{Center for Quantum Information and Control (CQuIC), Department of Physics and Astronomy, University of New Mexico, Albuquerque, New Mexico 87131, USA}

\date{\today}
\begin{abstract}
We study the generation of two-qudit entangling quantum logic gates using two techniques in quantum optimal control.  We take advantage of both continuous, Lie algebraic control and digital, Lie group control. In both cases, the key is access to a time-dependent Hamiltonian which can generate an arbitrary unitary matrix in the group SU($d^2$). We find efficient protocols for creating high-fidelity entangling gates. As a test of our theory, we study the case of qudits robustly encoded in nuclear spins of alkaline earth atoms and manipulated with magnetic and optical fields, with entangling interactions arising from the well-known Rydberg blockade. We applied this in a case study based on a $d=10$ dimensional qudit encoded in the $I=9/2$ nuclear spin in $^{87}$Sr, controlled through a combination of nuclear spin-resonance,  a tensor AC-Stark shift, and Rydberg dressing, which allows us to generate an arbitrary symmetric entangling two-qudit gate such as CPhase. Our techniques can be used to implement qudit entangling gates for any  $2\le d \le10$ encoded in the nuclear spin. We also studied how decoherence due to the finite lifetime of the Rydberg states affects the creation of the CPhase gate and found, through numerical optimization, a fidelity of $0.9985$, $0.9980$, $0.9942$, and $0.9800$ for $d=2$, $d=3$, $d=5$, and $d=7$  respectively. This provides a powerful platform to explore the various applications of quantum information processing of qudits including metrological enhancement with qudits, quantum simulation, universal quantum computation, and quantum error correction.

\end{abstract}
\maketitle

\section{Introduction} 
\label{sec:introduction}

In the standard paradigm of quantum information processing (QIP) one encodes information in qubits, the quantum analog of classical bits, by isolating two well-chosen energy levels of the system. 
In many platforms, one has access and control over multiple levels, which can enhance our ability to do QIP in a variety of ways~\cite{wang2020qudits,blok2021quantum,Gross2021,puri2020bias,PhysRevA.64.012310}. 
In particular, one can encode information in base-$d>2$ using $d$-level qudits~\cite{wang2020qudits}.
With a larger state space per subsystem, qudits offer potential advantages for quantum communication~ \cite{PhysRevLett.90.167906}, quantum algorithms~\cite{luo2014geometry,luo2014universal,li2013geometry,lu2020quantum}, and  topological quantum
systems~\cite{cui2015universal,cui2015universal1,bocharov2015improved} .
Quantum computation with qudits can also reduce circuit complexity and can be advantageous in a variety of NISQ-era applications~\cite{brylinski2002universal,lu2020quantum,luo2014universal,li2013geometry,
zobov2012implementation,weggemans2022solving,PhysRevLett.129.160501}.

Qudits may also provide significant advantages in quantum error correction and fault-tolerant quantum computation \cite{campbell2014,PhysRevA.83.032310,gottesman1998fault,campbell2012,Eliot2016}. { Of particular importance is reducing the physical resources needed to encode logical qubits.  In the standard paradigm, logical qubits are encoded in multiple physical qubits, such as in the well-known surface code~\cite{PhysRevA.86.032324}, which has a substantial overhead. An alternative approach is to encode a logical qubit in a single qudit.  This has been a powerful tool, e.g., in encoding a logical qubit in the multiple harmonic levels of a bosonic mode~\cite{PhysRevA.64.012310}, and has been theoretically considered in high dimensional spin qudits in atoms~\cite{Gross2021,omanakuttan2023multispin}, molecules~\cite{PhysRevX.10.031050}, and in solid state devices~\cite{gross2021hardware}.  Developing general methods for quantum control and entanglement of qudits would greatly expand the tools at our disposal.}

In the gate-based approach to quantum computation with qubits, a universal gate set consists of single-qubit gates that generate the group SU$(2)$ and one entangling two-qubit gate, such as CNOT ~\cite{divincenzo1995two}. 
This generalizes simply for qudits. 
The universal gate-set consists of the generators of single-qudit gates in SU($d$) and an entangling two-qudit gate \cite{muthukrishnan2000multivalued,zhou2003quantum,brennen2005criteria}. 
 Unlike qubits, where native Hamiltonians can be used to naturally implement the desired gate set, qudits require more complex protocols. 
 The gates that are necessary for the implementation of the universal gate set have been recently implemented for qudits in superconducting transmon~\cite{blok2021quantum,goss2022high,fischer2022towards} as well as in trapped ions~\cite{ringbauer2021universal,hrmo2022native} up to dimension $d=7$.  In these experiments, one implements qudit gates using constructive methods through a prescribed set of Givens rotations~\cite{brennen2005criteria,li2013decomposition}. 
 While there has been substantial progress, much work remains to be done to efficiently implement a high-fidelity universal qudit gate set. 

In this article, we study an alternative approach based
on quantum optimal control. 
Quantum optimal control was originally developed in NMR~\cite{vandersypen2005nmr} and for coherent control of chemistry~\cite{rabitz2000whither,shapiro2012quantum}, and has been extensively used in quantum information processing~\cite{koch2022quantum}. We consider both continuous Hamiltonian control (Lie algebraic) and digital gate-based control (Lie group). 
Quantum optimal control has been experimentally implemented in a wide range of platforms ranging from ion traps \cite{PhysRevA.82.012339}, neutral atoms~\cite{PhysRevA.74.022312,PhysRevA.90.032329,lysne2020small}, superconductors \cite{PhysRevLett.102.090401,goerz2014optimal}, and nitrogen vacancy (NV) centers \cite{waldherr2014quantum,scheuer2014precise}. Its use in  implementing single-qudit gates was demonstrated in the seminal experiments of Jessen~\cite{anderson2013unitary} with information encoded in the hyperfine states of cesium and studied in~\cite{omanakuttan2021quantum} for qudits encoded in the nuclear spin of alkaline-earth atoms. In this work, we extend these techniques to the implementation of entangling gates between two qudits. 
We study qudit entangling gates for any $k \leq d$ within the $d$-dimensional Hilbert space of each subsystem.


As a concrete example that demonstrates the power of the method, we present here an optimal control scheme to implement entangling gates in qudits encoded in the nuclear spin of $^{87}$Sr atoms. 
The nuclear spin is a good memory for use in quantum information processing given its weak coupling to the environment and resilience to other background noise~\cite{barnes2021assembly,PhysRevLett.101.170504,daley2011quantum}. 
{The ground state of the $^{87}$Sr is also studied in a recent paper as a possible candidate for qudit encoding with entangling interaction enabled by the Rydberg blockade \cite{zache2023fermion} }
Also, the recent significant achievements of quantum information processing using the Rydberg blockade~\cite{Levine2019,bluvstein2022quantum,graham2022multi}  make this an ideal platform for exploring quantum computation.  Using a combination of a tunable radio-frequency magnetic field and interactions that arise when atoms are excited to high-lying Rydberg states, the atomic qudit is fully controllable. We find that one can use quantum optimal control to implement  high-fidelity entangling qudit gates even in the presence of decoherence arising from the finite Rydberg-state lifetime. 

The remainder of this article is organized as follows. In Sec.~\ref{sec:controllability} we review the fundamentals of quantum control and define two approaches: Lie algebraic and Lie group theoretic protocols for the generation of any arbitrary qudit entangling gates. In Sec.~\ref{sec:Numerical Methods}, we study how control is achieved using  numerical optimization based on the well-known GRAPE algorithm~\cite{khaneja2005optimal} and obtain control waveforms using the Lie algebraic method. We also use a gradient-based approach to find a digital sequence of unitary maps that achieves the desired gate using a Lie group theoretic method. Finally, we study how decoherence affects the fidelity of these gates. We give conclusions and outlook of our approach in Sec.~\ref{sec:conclusions_and_future_work}.

\section{Controllability} 
\label{sec:controllability}
A complete universal gate set for qudits requires one entangling gate.
A standard choice is the CPhase gate, which is the generalization of CZ gate for qubits, defined 
\begin{equation}
\mathrm{CPhase}\ket{j}\ket{k}=\omega^{jk}\ket{j}\ket{k},
    \label{eq:CPhase_gate}
\end{equation}
where $\omega=\exp(2\pi i/d)$, the $d$-th primitive root of identity for a subsystem of dimension $d$. We can see that for $d=2$ we recover the CZ gate. 
This gate is locally equivalent to the qudit-analog of the CNOT gate, known as CSUM gate,
\begin{equation}
C_{\mathrm{SUM}}\ket{i}\ket{j}=\ket{i}\ket{i\oplus j (\text{mod } d)}
    \label{eq:CSUM_gate}
\end{equation}
by the Hadamard gate for qudits,  $H_d\ket{j}=\frac{1}{\sqrt{d}}\sum_i\omega^{ij}\ket{i}$. 
Previous works have studied how to implement these gates through a well-defined sequence of  maps generated by one-qudit and two-qudit Hamiltonians~\cite{brennen2005criteria,PhysRevA.62.052309,vlasov2002noncommutative,brylinski2002mathematics}. We study here the use of numerical optimization and the theory of optimal control.

\subsection{Lie algebraic approach }
\begin{figure*}
	\includegraphics[width=0.98\textwidth]{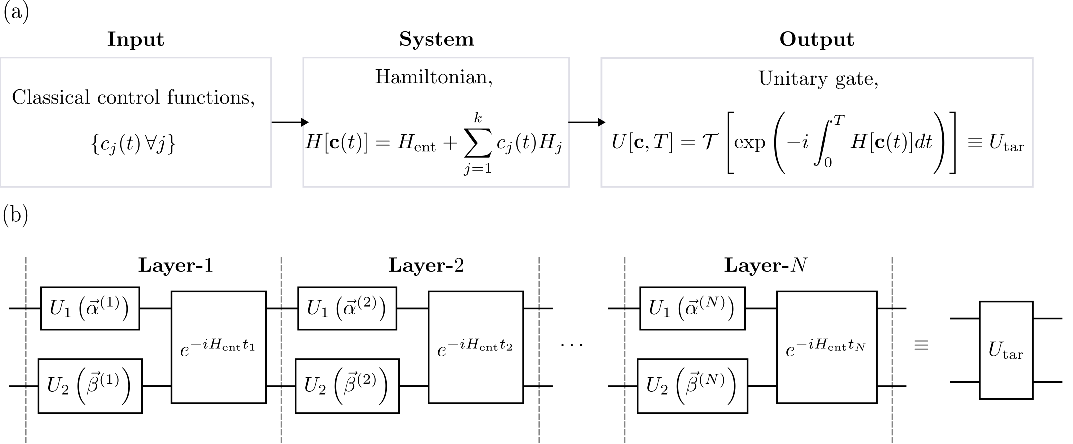}
\caption{ \textbf{Comparison of Lie algebra versus Lie group approach for quantum control.} (a) Schematic of the continuous-time Lie algebraic approach for quantum control. 
The physical systems are governed by the time-dependent Hamiltonian, $H[\mathbf{c}(t)] = H_{\mathrm{ent}} + \sum_{j=1}^{k} c_j(t) H_j$, here with a time-dependent entangling Hamiltonian, $H_{\mathrm{ent}}$. The time-dependent waveforms $\{c_j(t)\}$ are found through numerical optimization, and this defines the target unitary map of interest through the solution to the time-dependent Schr\"{o}dinger equation. 
(b) Schematic for a digital, Lie group approach to quantum control of entangling two-qudit gates. The target unitary is achieved through a discrete series of layers consisting of unitary maps from a given family. One layer of the scheme consists of single-qudit gates on each subsystem and an entangling interaction between them, applied for a given time $t_j$. Through numerical optimization, one finds the parameters of the local $SU($d$)$-gates and the entangling time $t_j$ in each layer.}
	\label{fig:approach_all}
\end{figure*}

In the Lie algebraic approach to quantum control, we consider a Hamiltonian of the form $H[\mathbf{c}(t)] = H_0 + \sum_{j=1}^{k} c_j(t) H_j$, where $\mathbf{c}(t)=\{c_j(t)\}$ is the set of time-dependent classical control waveforms, and $H_0$ is called the drift Hamiltonian. The system is said to be ``controllable" if the set of Hamiltonians, $\{H_0, H_1,H_2,\hdots, H_k\}$, are generators of the desired Lie algebra, e.g., $\mathfrak{su}{(d)}$. Then $\exists \hspace{0.1cm} \mathbf{c}(t)$ \hspace{0.2cm } such that $U[\mathbf{c},T]=\mathcal{T}\left[\exp\left(-i\int_0^T H[\mathbf{c}(t)]dt\right)\right]=U_{\mathrm{tar} }$ for any target unitary in desired Lie Group, e.g., $U_{\mathrm{tar}} \in \mathrm{SU}(d)$.
In addition, we require $T \ge T_*$, where $T_*$ is known as the ``quantum speed limit time," which sets the minimal time needed for the system to be fully controllable.


We consider here open-loop control determined by a well-defined Hamiltonian of the general form,
\begin{equation}
H(t)=H^{(1)}(t)+ H^{(2)}(t)+H_{\mathrm{ent}},
    \label{eq:entangling_qudit_Hamiltonian_1}
\end{equation} 
where $H^{(i)}(t)$ are time-dependent Hamiltonians acting on the individual subsystems, and $H_{\mathrm{ent}}$ is  the interaction that entangles them.  Here we include the time dependence in the Hamiltonian that acts on the individual system as these will be generally easier to implement experimentally. In this formulation, $H_{\mathrm{ent}}=H_0$, is the drift Hamiltonian.
However, one could in principle include time dependence in the entangling Hamiltonian as well and this may achieve faster gates.

\subsection{Lie group approach}

 In the digital, Lie group approach to quantum control, we consider a  family of unitary maps in the desired group  that are easily implementable, $U(\lambda_j)$, where $\{\lambda_j\}$ are the parameters that specify the unitary matrices at our disposal. The relevant Lie group of interest here is $\mathrm{SU}(d^2)$, the group of two-qudit unitary matrices in $d^2$ dimensions, where the overall phase is removed. The system is controllable if  $\forall U_{\mathrm{tar}} \in \mathrm{SU}(d^2)$, $\exists \{\lambda_i\}$ such that  $\prod_{j=1}^{k}U(\lambda_j)=U_{\mathrm{tar} }$. Similar to the Lie algebraic quantum control approach, the goal is to find  $\{\lambda_j\}$ through numeric optimization, e.g., via gradient-based methods.

For the case of two-qudit gates, a controllable Lie group structure is given as,
\begin{equation}
    U_{\lambda_j}=U_{\mathrm{ent}}*(U_1\otimes U_2),
    \label{eq:Lie group_approach}
\end{equation}
where $U_{1,2}\in \mathrm{SU}(d) $ and  $U_{\mathrm{ent}}=\exp(-i H_{\mathrm{ent}}t) \notin \mathrm{SU}(d)\otimes \mathrm{SU}(d)$.
Thus, we can achieve the target gate to the desired fidelity by intertwining a sequence of local $\mathrm{SU}(d)$ gates and the available entangling interaction  in alternating layers of single qudit gates and entangling gates, as shown in Fig.~\ref{fig:approach_all}(b).
This approach is similar to the construction based on Givens rotation~\cite{ringbauer2021universal}.
Here, the possibility of accessing arbitrary local $\mathrm{SU}(d)$ gates makes this protocol very powerful. A schematic comparison of both these approaches is shown in Fig.~\ref{fig:approach_all}. 

\subsection{Physical Platform: Rydberg atoms }
To make these ideas concrete, we consider the implementation of entangling gates in neutral atoms using the strong van~der~Waals interactions between atoms in high-lying Rydberg states. 
We use the Rydberg dressing paradigm in which one adiabatically superposes the Rydberg state into the ground states to introduce interactions between dressed ground states ~\cite{johnson2010interactions, keating2015robust, jau2016entangling, zeiher2016many, zeiher2017coherent, borish2020transverse}. 
Rydberg dressing has been studied with multiple applications including  the  dynamics of interacting spin models ~\cite{zeiher2016many, zeiher2017coherent, borish2020transverse} as well as to prepare metrologically-useful states ~\cite{kaubruegger2019variational}.
Entanglement between neutral atoms via Rydberg dressing has been theoretically proposed for creating qubit entangling gates ~\cite{keating2015robust, mitra2020robust,mitra2022practical} and experimentally implemented ~\cite{jau2016entangling,martin2021molmer,schine2021long}. 

We study here encoding a qudit in the spin of $^{87}$Sr. In the ground state there is neither orbital nor spin angular momentum in the electrons, $J=0$, and only nuclear spin, $I=9/2$, giving us ten possible levels in which to encode our qudit, labeled from  $\ket{0}=\ket{m_I=9/2}, \ket{1}=\ket{m_I=7/2},\cdots ,\ket{9}=\ket{m_I=-9/2}$. The nuclear spin is highly isolated from the environment and thus serves as a robust memory for quantum information. In \cite{omanakuttan2021quantum} we studied how we could implement single qudit gates in these systems through a combination of a laser-induced tensor light-shift and radio frequency (rf)-induced Larmor precession. We generalize to the two-qudit case here. 

To implement entangling two-qudit control, we will make use of the excitation to the $5sns\hspace{0.1cm}^3 S_1$ Rydberg series from the metastable $5s5p\hspace{0.1cm}^3 P_J$ first excited states. The choice of energy levels depends on practical considerations. 
By choosing the metastable state $5s5p\hspace{0.1cm}^3P_2$, the total electron angular momentum gives rise to a large magnetic dipole moment, which substantially increases the speed of gates, but this comes at the cost of increased sensitivity to background magnetic fields and residual tensor light shifts from the trapping lasers. By choosing the metastable $^3{P}_0$ clock state, one retains the robustness of having $J=0$ at the cost of much slower gates, since the magnetic coupling is now solely to the nuclear spin. We consider here coherently transferring qudits from the $^1{S}_0$ ground state to the $F=9/2$ state hyperfine states of the  $^3{P}_2$ manifold, which provides for faster and more flexible control \cite{trautmann20221}, putting technical noise aside.  

To achieve the entangling interaction, we consider Rydberg dressing, generalizing the mechanism discussed in~\cite{jau2016entangling, keating2015robust,  mitra2020robust}. 
The AC Stark shift (light shift) associated with a dressed state when a laser is tuned near a Rydberg resonance is modified for two atoms because of the Rydberg blockade. 
The deficit between the two-atom light shift and twice the one-atom light shift determines the entangling  energy~\cite{keating2015robust}. 
For the case of qudits, the same physics holds, but now with a multilevel structure and a spectrum of entangling energies. 
When the spectrum is nonlinear, the system is controllable.

\begin{figure*}
    \includegraphics[width=0.98\textwidth]{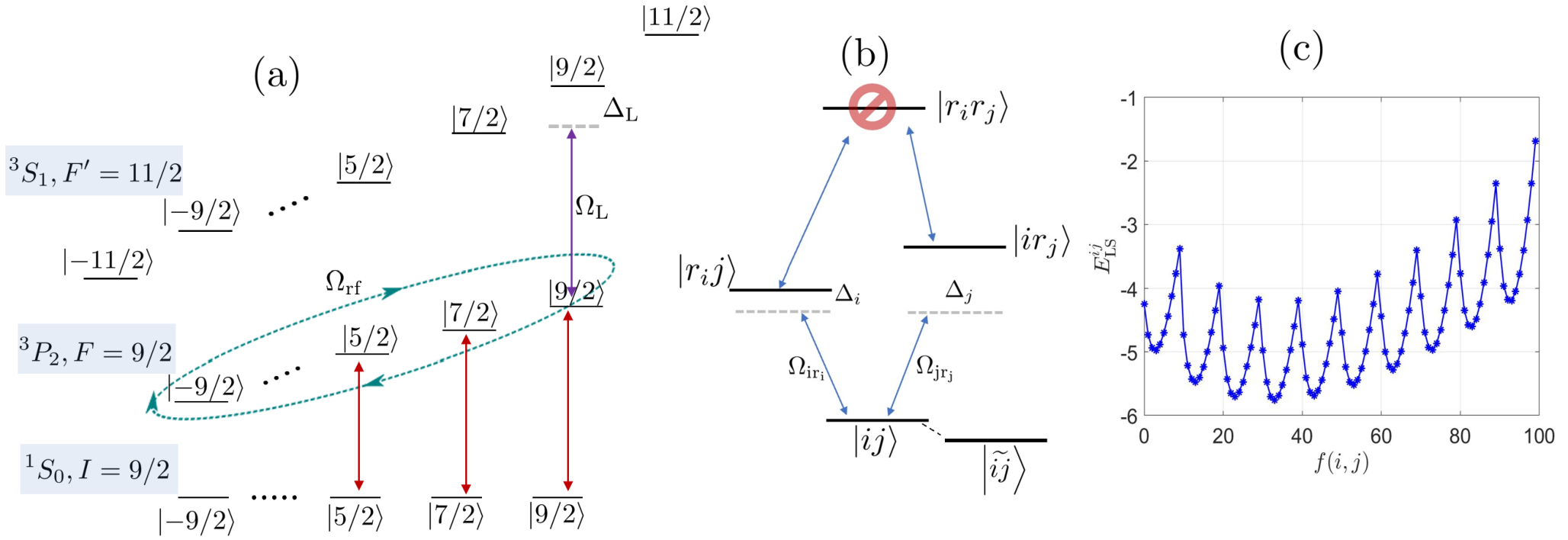}
\caption{\textbf{Schematic for designing two-qudit entangling interactions in $^{87}$Sr neutral atoms.} (a) A $k\le d$-dimensional qudit is encoded in memory in the nuclear spin with $d=10$ magnetic sublevels in the electronic ground state $(5s^2) \;^1S_0$. When the gate is to be performed, the $k$ levels (here $k=3$) are transferred coherently to the  metastable clock states $ (5s5p)\; ^3P_2, F=11/2$ in the presence of a bias magnetic field.  The system becomes controllable  by adiabatically dressing the $^3P_2$ with Rydberg character through the application of a near-resonant laser with Rabi frequency $\Omega_{\mathrm{L}}$ and detuning $\Delta_{\mathrm{L}}$ with respect to the hyperfine manifold $(5sns)\; ^3S_1, F'=9/2$ in the Rydberg series. 
Control is then achieved through the application of a phase-modulated rf-field with Rabi rate $\Omega_{\mathrm{rf}}$ which acts on the dressed states to generate a nonlinear Larmor precession. The entanglement arises due to the Rydberg blockade. 
The coupling of the state of two qudits for a perfect blockade as depicted in (b), where $i$ is a state from the first qudit and $j$ is from the second qudit, excited by two Rabi frequencies and detunings determined by the Clebsch-Gordan coefficients and Zeeman shifts. The state $\ket{ij} \to  \ket{\widetilde{ij}}$ is the dressed state given in Eq. \eqref{eq:dressed_states_qudits}.
The spectrum of eigenvalues of the entangling Hamiltonian Eq.\eqref{eq:entangling_Hamiltonian} is given in (c) as a function of $i$ and $j$ where the function chosen is $f(i,j)=10i+j; 0\leq i,j<10$. The spectrum indicates 10 parabolas, where each parabola corresponds to the effect of a single state in the first atom sees due to all the states in the second atom. This nonlinear spectrum arises through a combination of the tensor AC Stark shift and the Rydberg blockade, making the system controllable, allowing us to implement any symmetric two qudit gate in this system of interest. }
	\label{fig:set_up_figure}
\end{figure*}

 Fig.~\ref{fig:set_up_figure} depicts the basic scheme. Those levels of the qudit that we chose to participate in the gate are  excited from the ground  ${^1S_0}$ to the first excited ${^3P_2}$ state. 
 The Rydberg states in $^{87}$Sr have well-resolved hyperfine splitting. We consider UV dressing laser near the resonance between the ${^3P_2}$, $F=9/2$ hyperfine manifold and the ${^3S_1}$, $F'=11/2$ Rydberg hyperfine states. 
 In the presence of a bias magnetic field, due to the difference in the g-factors, the two manifolds will be differently Zeeman shifted. The different magnetic sublevels that define the qudit will thus be differently detuned to the Rydberg magnetic sublevels. Due to this and the Clebsch-Gordan coefficients associated with the different transitions, each sublevel will be differently dressed (equivalently, there is a tensor light shift). When two atoms are dressed, the effect of the Rydberg blockade modifies the spectrum as discussed above. 
 
An example of two sublevels (one from each atom) is shown in Fig.~\ref{fig:set_up_figure}(b). 
Diagonializing this atom-laser Hamiltonian under the approximation of a perfect Rydberg blockade yields the representation
\begin{equation}
    H_{\mathrm{ent}}=\sum_{ij} E^{ ij}\ket{\widetilde{ij}}\bra{\widetilde{ij}},
    \label{eq:entangling_Hamiltonian}
\end{equation}
where the tilde indicates dressed states,
\begin{equation}
\ket{\widetilde{ij}}=C_{ij}\ket{ij}+C_{r_ij}\ket{r_ij}+C_{ir_j}\ket{ir_j},
    \label{eq:dressed_states_qudits}
\end{equation}
and $E^{ij}$ are the light shifts originating from these interactions.
The spectrum of the entangling Hamiltonian shown in Fig.~\ref{fig:set_up_figure}(c) gives us insight into the controllability of the system. In the chosen order, the spectrum reveals the structure of $10$ quadratic potentials arising from a combination of the tensor light shift and Rydberg blockade. This nonlinearity makes the Hamiltonian controllable; further details are discussed in Appendix~(\ref{sec:controllability_of_the_Hamiltonian}).

  The time-dependent Hamiltonian necessary for the Lie algebraic control can be chosen as phase-modulated Larmor precession, $H_{\mathrm{mag}} = -\boldsymbol{\mu}\cdot \mathbf{B}(t)$, with $\boldsymbol{\mu} = g_F \mu_B \mathbf{F}$ the magnetic dipole vector operator, and where $\mathbf{B}(t) = B_\parallel \mathbf{e}_z + B_T \Re\left[(\mathbf{e}_x + i \mathbf{e}_y) \mathbf{e}^{-i\left(\omega_\text{rf} t +\phi(t)\right)}\right]$. 
Defining the auxilary subspace, $a$, for the levels in hyperfine manifold $\{5s 5p\hspace{0.1 cm}^3P_2,\hspace{0.1 cm} F=9/2\}$ and the subspace, $r$, for the levels $\{5s ns \hspace{0.1 cm}^3S_1,\hspace{0.1 cm} F'=11/2\}$ in the Rydberg hyperfine manifold, we have  $g_F(r)/g_F(a)\approx 2$. 
Thus defining the Zeeman shift $\omega_0= g_F(a)B_\parallel $, the Larmor precession frequency $\Omega_{\mathrm{rf}}= g_F(a)B_T $, and choosing rf drive on resonance in the $a$-manifold, $\omega_{\mathrm{rf}}=\omega_0$, in the co-rotating frame at $\omega_0$, the Hamiltonian is
\begin{equation}
\begin{aligned}
H_{\mathrm{mag}}^{(a)}(t)&=\Omega_{\mathrm{rf}} \left[\cos \phi(t) F_x^{a} +\sin \phi(t) F_y^{a}\right],\\
H_{\mathrm{mag}}^{(r)}(t)&=2\Omega_{\mathrm{rf}} \left[\cos \phi(t) F_x^{r} +\sin \phi(t) F_y^{r}\right]+\omega_{0}F_z^{r},
\end{aligned}
    \label{eq:Hamiltonian_frame}
\end{equation}
where $F_i^{a}, F_i^{r}$ are the spin angular momentum operators in the respective subspaces along axis $i \in \{x, y, z\}$.



As the $H_\mathrm{mag}$ acts on the laser-dressed states defined in Eq.~(\ref{eq:dressed_states_qudits}), which are superpositions of $a$ and $r$ states that have different $g$-factors, one needs to find the action of the magnetic interaction in the dressed basis. Due to the nonlinearity, the action of the rf-magnetic driving on the dressed states is no longer simple Larmor precession. Considering a global rf-magnetic interaction, the $H_\mathrm{mag}$ acts on both qudits as
\begin{widetext}
\begin{equation}
\begin{aligned}
    \left(H_\mathrm{mag}(t)\otimes \mathds{1}+\mathds{1}\otimes H_\mathrm{mag}(t)]\right)\ket{\widetilde{ij}}&=C_{ij}\left[H_{\mathrm{mag}}^{(a)}(t)\otimes H_{\mathrm{mag}}^{(a)}(t)\right] \ket{ij}+C_{r_ij}\left[H_{\mathrm{mag}}^{(r)}(t)\otimes H_{\mathrm{mag}}^{(a)}(t)\right] \ket{r_ij}\\
    &+C_{ir_j}\left[H_{\mathrm{mag}}^{(a)}(t)\otimes H_{\mathrm{mag}}^{(r)}(t)\right] \ket{ir_j}.
\end{aligned}
\end{equation}
\end{widetext}
Thus in the dressed basis, the Hamiltonian is $H(t)=\widetilde{H}\left[\phi(t)\right]+H_{\mathrm{ent}}$,
where the action of the magnetic field in the dressed basis is given by the Hamiltonian,
\begin{equation}
\begin{aligned}
&
\widetilde{H}\left[\phi(t)\right]
\\ &
=\sum_{i,j,k,l}\bra{\widetilde{ij}}H_\mathrm{mag}(t)\otimes \mathds{1}+\mathds{1}\otimes H_\mathrm{mag}(t)]\ket{\widetilde{kl}} \ketbra{\widetilde{ij}}{\widetilde{kl}}.
    \label{eq:dressed_rotation}
\end{aligned}
\end{equation}
By modulating the phase $\phi(t)$ one can generate any target unitary gate.


\section{Numerical Methods} 
\label{sec:Numerical Methods}
We consider encoding a $k$-dimensional qudit in the $d=10$ dimensional Hilbert space associated with $10$ magnetic sublevels of the nuclear spin of $^{87}$Sr. To implement gates based on optimal control for $k<10$, we use techniques based on the structure of  partial isometries. A partial isometry of dimension $k \le d$ in a physical system of dimension $d$ is defined as,
\begin{equation}
V_\mathrm{{tar}}=\sum_{i=1}^{k}\ket{f_i}\bra{e_i}
\label{eq:Partial_isometry_definition}
\end{equation}
where $\{\ket{e_i}\},\{\ket{f_i}\}$ are two orthonormal bases for the qudit. The unitary of maps of interest then has the form, 
\begin{eqnarray}
U_{\mathrm{tar}}=V_{\mathrm{tar}}+V_{\perp},
\end{eqnarray}
where $V_{\perp}$ acts on the orthogonal subspace, with dimension $d-k$. To find the  control waveform, one then optimizes the fidelity between the target isometry and the isometry generated using quantum control~\cite{pedersen2007fidelity}
\begin{eqnarray}
\mathcal{F}_V[\bm{c},T]&=&\left|\Tr\left(V^{\dagger}_{\text{tar}}V[\bm{c},T]\right)\right|^2/k^2.
\label{eq:fidelity}
\end{eqnarray} 


\subsection{Numerical results for Lie algebraic approach}

As discussed in Sec.~\ref{sec:controllability}c, one can implement an arbitrary entangling gate through a combination of Rydberg dressing and phase-modulated Larmor precession driven by rf-fields. 
Because our control Hamiltonian is symmetric with respect to the exchange of the qudits, we consider here symmetric gates, with global control. We seek, through numerical optimization, the  time-dependent rf-phase, $\phi(t)$. 
To achieve this we employ the well-known GRAPE algorithm~\cite{khaneja2005optimal}. 
To implement GRAPE, we discretize the control waveform, $\phi(t)$, and  numerically maximize the fidelity by gradient ascent.  We choose here a piecewise constant parameterization (as in ~\cite{anderson2013unitary}) and write the control waveform as a vector $\mathbf{c} = \{\phi(t_j)/\pi  \;|\; j=1,\dots,n\}$ where $t=j\Delta t$ and $n=T/\Delta t$. The waveform is thus a series of square rf-pulses with constant amplitude and phase over the duration $\Delta t$. 

The minimum number of elements in the control vector $\mathbf{c}$ is determined by the number of parameters needed to specify the target isometry. 
A $K$-dimensional partial isometry is defined by the $K$ columns in a $D \times D$-dimensional unitary matrix.
Hence, to find the number of free parameters for a $K$-dimensional isometry one can count the number of parameters needed to specify $K$ orthonormal vectors uniquely in a  $D$-dimensional vector space. This is given by 
\begin{equation}
    \begin{aligned}
        n_{\mathrm{min}}(K,D)=&\sum_{j=1}^{K}2(D-j)-1+K-1\\
        =&2\left[KD-\frac{K(K+1)}{2}\right]+K-1\\
        =&2KD -K^2-1,
    \end{aligned}
    \label{eq:number_of_free_parameters}
\end{equation}
where in the first line, we subtracted one from the parameter count in since the overall phase of the isometry is neglected. Eq. \eqref{eq:number_of_free_parameters} recovers  well-known limits. When  $K=1$ and $D=d$,  $n_{\mathrm{min}}=2d-2$, which is the number of free parameters needed to specify a pure state in a $d$-dimensional Hilbert space. 
When $K=D=d$, $n_{\mathrm{min}}=d^2-1$, which is the number of free parameters needed to specify a special unitary map in $d$-dimensions.

In the Lie algebraic protocol for designing entangling gates, the  control Hamiltonian, as well as the target unitary matrices, are symmetric under the exchange of qudits.
In this case, one can work in the symmetric subspace for two qudits. Using the hook length formula~\cite{frame1954hook}, the dimension of the symmetric subspace of the total vector space and isometry is, 
\begin{equation}
D=\frac{d(d+1)}{2}, K=\frac{k(k+1)}{2}.
\label{eq:effective_parameters_unitary}
\end{equation}
Thus, using Eq.~(\ref{eq:number_of_free_parameters}), we find the  number of free parameters required for the two-qubit entangling unitary given in Table~\ref{tab:Lie_algebra_parameters}. 

\begin{table}
 \begin{tabular}{ |c|c| } 
 \hline
 $k$ & $n_{\mathrm{min}}(K,D)$ \\
\hline 
 2 & 320 \\
 \hline
3& 623  \\
 \hline
5 &  1424\\
\hline
7& 2295\\
\hline
\end{tabular}
    \caption{The minimum number of parameters required for  encoding a partial isometry of dimension $k$ in the $d=10$ dimensional Hilbert space according to Eq.~\eqref{eq:number_of_free_parameters} for the prime dimensions $k\leq 10$ with $K$ and $D$ given by Eq.~\eqref{eq:effective_parameters_unitary}}
    \label{tab:Lie_algebra_parameters}
    
\end{table}

\begin{figure*}
\centering
\includegraphics[width=0.98\textwidth]{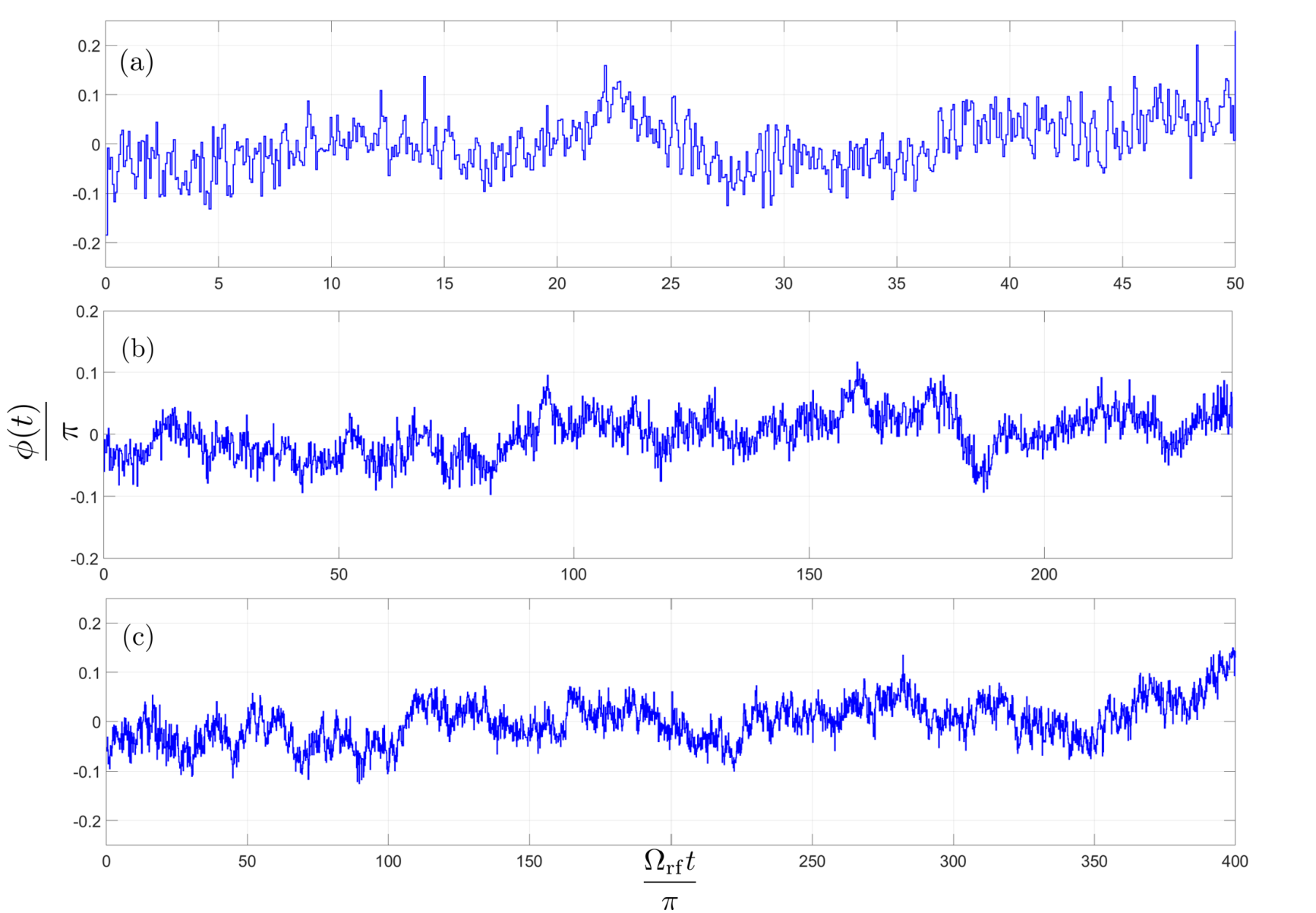}
    \caption{\textbf{Waveforms of the CPhase gate.} 
    Quantum control is achieved by modulating the phase of an rf-field as a function of time, $\phi(t)$. We parameterize this by a piecewise constant waveform. The figure shows proof-of-principle examples of $\phi(t)$ that generate the CPhase gate, optimized using the GRAPE algorithm for different qudit dimensions. (a) The case of the $d=3$ for a total time of $T=50 \pi/\Omega_{\mathrm{rf}}$ with $700$ piecewise constant steps. 
    (b) The case of the $d=5$ for a total time of $T=240 \pi/\Omega_{\mathrm{rf}}$ with $1600$  piecewise constant steps. (c)  The case of the $d=7$ for a total time of $T=400 \pi/\Omega_{\mathrm{rf}}$ with $2500$ piecewise constant steps. 
    For all of these calculations, the rf-field is on resonance with the Zeeman splitting $\omega_{\mathrm{rf}}=\omega_0$ and we choose the rf-Larmor frequency $\Omega_{\mathrm{rf}}=\omega_{\mathrm{rf}}$. Control is achieved by Rydberg dressing with laser Rabi frequency $\Omega_{\mathrm{L}}=6 \Omega_{\mathrm{rf}}$.}
    \label{fig:Qudits_simulations}
\end{figure*}

Proof-of-principle numerical examples of  waveforms that generate the CPhase gate are given in Fig.~\ref{fig:Qudits_simulations}. 
The figure gives the $\phi(t)$ as a piecewise constant function of time, obtained using the GRAPE algorithm. We consider prime-dimensional qudits, the cases of most interest in quantum algorithms. Fig.~\ref{fig:Qudits_simulations}(a) shows the case of the $k=3$, a qutrit encoded in $d=10$. The total time is $T=50 \pi/\Omega_{\mathrm{rf}}$, which is divided into $700$ intervals for the quantum control. 
Fig.~\ref{fig:Qudits_simulations}(b) shows an example waveform for the case of $k=5$. Here, the total time is $T=240 \pi/\Omega_{\mathrm{rf}}$,  divided into $1600$ intervals. 
Similarly, Fig.~\ref{fig:Qudits_simulations}(c) shows the case of $k=7$ in our $d=10$ level system. 
The total time is $T=400 \pi/\Omega_{\mathrm{rf}}$, divided into 2500 time intervals. This controllable Hamiltonian can also be used to generate other two-qudit gates. The qudit generalization of the M$\o$lmer-S$\o$rensen gate, as is given in the Appendix~\ref{sec:Molmer_sorenson_gate}.

The waveforms found here are a proof-of-principle set of square pulses and are not intended to be taken as the best choice for experimental implementation. In practice, one can design and optimize for much smoother waveforms using well-known techniques by imposing additional constraints on bandwidth and slew rate. Alternatively, one can optimize in the Fourier domain or in any other complete basis of functions using the techniques of gradient optimization of analytic controls (GOAT) \cite{machnes2015gradient}.

 \subsection{Numerical results for Lie group approach}
In the Lie group control protocol discussed in Sec.\ref{sec:controllability}c  we parameterize the target unitary map as 
\begin{equation}
    \begin{aligned}
     U_{\mathrm{tar}}&=\prod_{j}U_{\lambda_j},\\
     &=\prod_{j} e^{-i H_{\mathrm{ent}}t_j}\,
    U_1(\vec{\alpha}^{(j)})\otimes U_2(\vec{\beta}^{(j)}).
\end{aligned}
\label{eq:Lie_group_approach_1}
\end{equation}
The control parameters $\{\lambda_i\}$ consist of the set of times $\{t_i\}$ and the $2(d^2-1)$ parameters $\vec{\alpha}^{(j)}$, $\vec{\beta}^{(j)}$, which  specify each of the local $\mathrm{SU}(d)$ unitary maps.  We can parameterize these according to
\begin{equation}
    \begin{aligned}
     U_i(\vec{\alpha}^{(j)})=\exp(-i\sum_{i=1}^{d^2-1}\alpha_i^{(j)}\Lambda_i),
    \end{aligned}
\end{equation}
where $\Lambda$ is the generalized Gell-Mann matrices that span the Lie algebra $\mathfrak{su}(d)$. The matrices can be categorized as,
\begin{equation}
    \begin{aligned}
    \text{symmetric: }&\Lambda_{jk}^{x}=\ketbra{j}{k}+\ketbra{k}{j},\\
    \text{anti-symmetric: }&\Lambda_{jk}^{y}=-i\ketbra{j}{k}+i\ketbra{k}{j},\\
    \text{diagonal: }&\Lambda_{l}^z=\sum_{j=1}^{l}\ketbra{j}{j}-l\ketbra{l+1}{l+1}.
    \end{aligned}
\end{equation}
The task of the numerical optimization, thus, is to find the set of times of the entangling interaction $\{t_j\}$, and the expansion coefficients of the Gell-Mann matrices $\{\alpha_i^{(j)}\}$ and  $\{\beta_i^{(j)}\}$. We denote this whole set of parameter as $\{\lambda_j\}=\{t_j,\vec{\alpha}^{(j)},\vec{\beta}^{(j)}\}$.

We define one layer of the control as consisting of a pair of local SU$(d)$ gates followed by the entangling Hamiltonian for a time $t_j$. 
The total number of free parameters for a CPhase gate is $d^2(d^2+1)/2$, as follows from Eq.~(\ref{eq:effective_parameters_unitary}) for a symmetric gate in $\mathrm{SU}(d^2)$.
Thus, the minimum number of layers  required to obtain the CPhase gate is given by
\begin{equation}
\begin{aligned}
N_{\text{min}}\left(2(d^2-1)+1\right)=&\frac{d^2(d^2+1)}{2}\\
N_{\text{min}}=&\frac{d^2(d^2+1)}{2(2d^2-1)}.
\end{aligned}
\label{eq:minimum_number_Lie_group}
\end{equation}
  The numerical results for the minimum number of layers needed in the system are given in Table~\ref{tab:Lie_group_asymmetric} for the cases of $d=3,5,$ and $d=7$. In practice, we find that one needs more than this minimum number of layers to implement the target unitary gate with high fidelity. This improves the optimization landscape for gradientascent~\cite{larocca2018quantum}.

 For our case under study, we choose the same entangling Hamiltonian as we used in the Lie algebraic approach given in Eq. (\ref{eq:entangling_Hamiltonian}). However, unlike that approach, we interleave the entangling interaction with local single-qudit SU($d$) gates. Implementation of this requires another layer of optimization. As we do not have access to native Hamiltonians proportional to the Gell-Mann matrices, to implement local qudit gates we can employ local SU($d$) optimal control~\cite{omanakuttan2021quantum}. From a practical perspective, this might be implemented directly in the $^3 P_2$ manifold, either through a combination of tensor-light shift and rf-driven Larmor precession similar to~\cite{omanakuttan2021quantum}, or alternatively through a combination of microwave-driven Rabi oscillations between different hyperfine levels in $^3 P_2$ and rf-driven Larmor procession as in~\cite{anderson2013unitary}. In either case, optimal control can be used to find the relevant experimental waveform that generates the desired local $\mathrm{SU}(d)$ gates.

\begin{table}
 \begin{tabular}{ |c|c|c|c| } 
 \hline
 $d$ & $N_{\mathrm{min}}$ & $N_{\text{local}}$  &$N_{\text{global}}$\\
\hline 
 3 & 3 &6 & 7\\
 \hline
5 & 7 & 10 & 12\\
 \hline
7 & 13 & 14 &15\\
\hline
\end{tabular}
    \caption{The number of layers of primitive gates in the Lie group approach required to achieve the CPhase gate. The theoretical minimum is $N_{\rm{min}}$ according to Eq.~\eqref{eq:minimum_number_Lie_group}. If we allow locally addressable single qudit gates, the number of layers required is $N_{\rm{local}}$. If we have only global control but allow for a sign change in the entangling Hamiltonian, the number of layers required is $N_{\rm{global}}$}
    \label{tab:Lie_group_asymmetric}
    
\end{table}

In this analysis, we included locally addressable control on each qudit. Though the CPhase gate is symmetric under exchange, we find that this symmetry breaking is necessary for effective optimization of this parameterization, similar to that seen in~\cite{PhysRevResearch.3.023092}.
An alternative protocol is to employ symmetric global control of the local unitaries, $\vec{\alpha}^{(j)}=\vec{\beta}^{(j)}$, but to reverse the sign of the entangling Hamiltonian $H_{\mathrm{ent}}\to - H_{\mathrm{ent}}$ in alternating layers. This allows for effective optimization, and the corresponding result is given in Table~(\ref{tab:Lie_group_asymmetric}).

\subsection{Decoherence}
In a closed quantum system, quantum optimal control employing either the Lie algebraic or the Lie group approaches can be used in principle to implement any qudit entangling gate to any desired fidelity. In our numerical optimization, we took the target infidelity to be $10^{-3}$. In the  absence of decoherence, we could achieve that target in a reasonable time for $d \le 5$. For $d=7$, more time is required. However,  the fundamentally achievable fidelity is limited by decoherence associated with the particular physical platform. For the system at hand, decoherence occurs due to the finite lifetime of the Rydberg states, which predominantly leads to leakage and loss outside the computational basis. In that case, we can model the gate as generated by a non-Hermitian effective Hamiltonian, $H_\mathrm{eff}[c(t)]$, where the Hermitian part is the control Hamiltonian and the anti-Hermitian represents decay out of the Rydberg states. The fidelity of interest is given by
\begin{equation}
\mathcal{F}_V[\bm{c},T]=\left|\Tr\left(V^{\dagger}_{\text{tar}}V_{\mathrm{eff}}[\bm{c},T]\right)\right|^2/d^2,
\label{eq:fidelity_1}
\end{equation} 
where  $V_{\mathrm{eff}}[\mathbf{c},T]=\mathcal{T}\left[\exp\left(-i\int_0^T H_{\mathrm{eff}}[\mathbf{c}(t)]dt\right)\right]$. Here  the decay amplitude from a dressed state is $\gamma_{\mathrm{decay}}^{ij}=|C_{r_ij}|^2\Gamma_{r_i}+|C_{ir_j}|^2\Gamma_{r_j}$, which in turn gives the effective Hamiltonian as 
\begin{equation}
     H_{\mathrm{ent}}^{\mathrm{eff}}=\sum_{ij} \left(E_{\mathrm{LS}2}^{ij}-i \gamma_{\mathrm{decay}}^{ij}/2\right)\ket{\widetilde{ij}}\bra{\widetilde{ij}}.
     \label{eq:effective_entangling_interaction}
\end{equation}

\begin{figure}
\centering
	\includegraphics[width=0.49\textwidth]{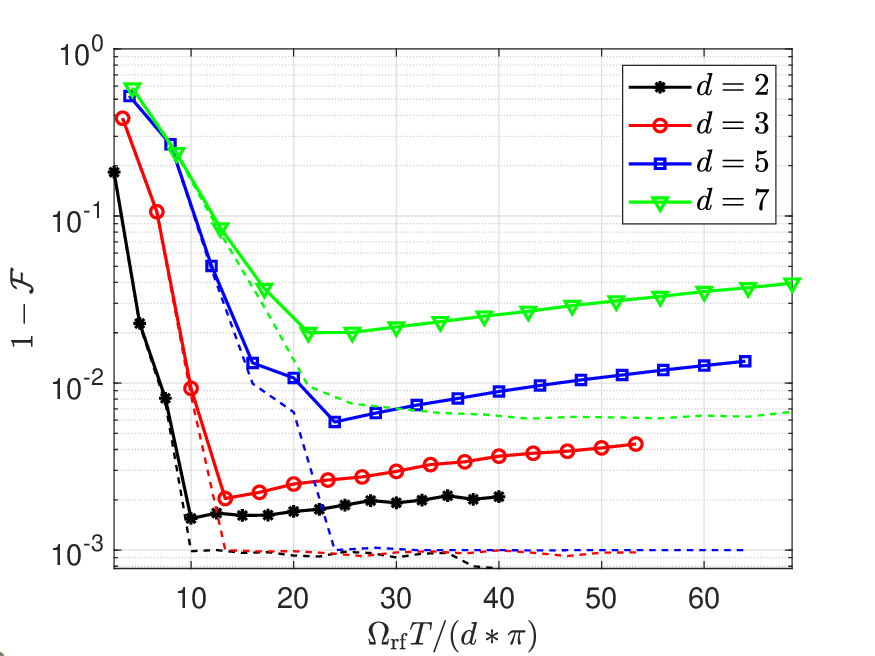}
	\caption{\textbf{Infidelity as a function of time.} 
	Simulated infidelity with and without decoherence as a function of control time divided by the dimension $d$ for CPhase gate with different prime dimensions with $d\leq 10$, as found using  Lie algebraic quantum control and the GRAPE algorithm. Decoherence  due to Rydberg decay outside the computational basis is included through an imaginary part of the Hamiltonian. We take the Rydberg lifetime to be $140 \mu$s and choose the rf-Larmor frequency to be $\Omega_{\mathrm{rf}}/2\pi=10$ MHz. In the absence of decoherence (dashed lines), for a time greater than the ``quantum speed limit" (the time required to obtain ideal fidelity) we achieve a minimal error (infidelity) of $10^{-3}$ due to our threshold in the numerics for $d\le 5$. This speed-limit time increases as we increase the qudit dimension, which in turn results in an increased decay in maximum fidelity.	For the CPhase gate, we obtain a fidelity of  $0.9985$, $0.9980$, $0.9942$, and $0.9800$ for $d=2, d=3, d=5, \text{ and } d=7$ respectively. For all of these calculations, we have taken the dressing laser Rabi frequency to be $\Omega_{\mathrm{L}}=6 \Omega_{\mathrm{rf}}$ and  the lifetime of the Rydberg states to be $140\mu$s.}
	\label{fig:Fidelity figure 1}
\end{figure}

With this model for decoherence in hand, the numerical results for the Lie algebraic approach are  given in Fig.~\ref{fig:Fidelity figure 1}, which shows the infidelity as a function of time for a CPhase gate for different dimension isometries. We focus here on the case of the prime dimensional qudits. In contrast to closed-system control, in the presence of decoherence, infidelity decreases at first and then increases. This is due to the fact there is an optimal time of evolution, larger than the quantum speed limit, but not too large when compared to the coherence time of the system. As expected, one needs more time as the qudit dimension increases, which in turn results in an increase in the minimum infidelity one could achieve in each of these cases as shown in Fig.~\ref{fig:Fidelity figure 1}. We obtain a maximum fidelity of 0.9985, 0.9980, 0.9942, and 0.9800 for $d=2$,  $d=3$, $d=5$, and $d=7$ respectively for the CPhase gate. Note, the values of fidelity for different dimensional qudits should be considered in the context of a particular application. For example, the threshold for fault tolerance for qudits, in general, is larger for larger $d$~\cite{anwar2014fast,watson2015fast}. For the particular scheme considered in~\cite{anwar2014fast}, the threshold for $d=2$, $d=3$, $d=5$, and $d=7$ are close to $0.008$, $0.012$, $0.0135$, and $0.015$ respectively. Hence, the proof-of-principle fidelity obtained here is promising and can be further optimized.

In the Lie group approach, we can use the effective Hamiltonian to describe the evolution when the Rydberg dressing is employed. In this case, we have,
\begin{equation}
    \begin{aligned}
     U_{\mathrm{tar}}^{\mathrm{eff}}&=\prod_{j}U_{\lambda_j},\\
     &=\prod_{j} e^{-i H_{\mathrm{ent}}^{\mathrm{eff}}t_j} U_1(\vec{\alpha}^{(j)})\otimes U_2(\vec{\beta}^{(j)}).\\
\end{aligned}
\label{eq:Lie_group_approach_eff}
\end{equation}
 We neglect here any decoherence associated with the local SU($d$) gates. Thus the fidelity including the decoherence effects is given as,
\begin{equation}
\mathcal{F}_{\mathrm{eff}}=\left|\Tr\left(U^{\dagger}_{\text{tar}}U_{\mathrm{tar}}^{\mathrm{eff}}\right)\right|^2/d^2,
\label{eq:Lie_group_fidelity}
\end{equation}
\begin{figure}
\centering
	\includegraphics[width=0.5\textwidth]{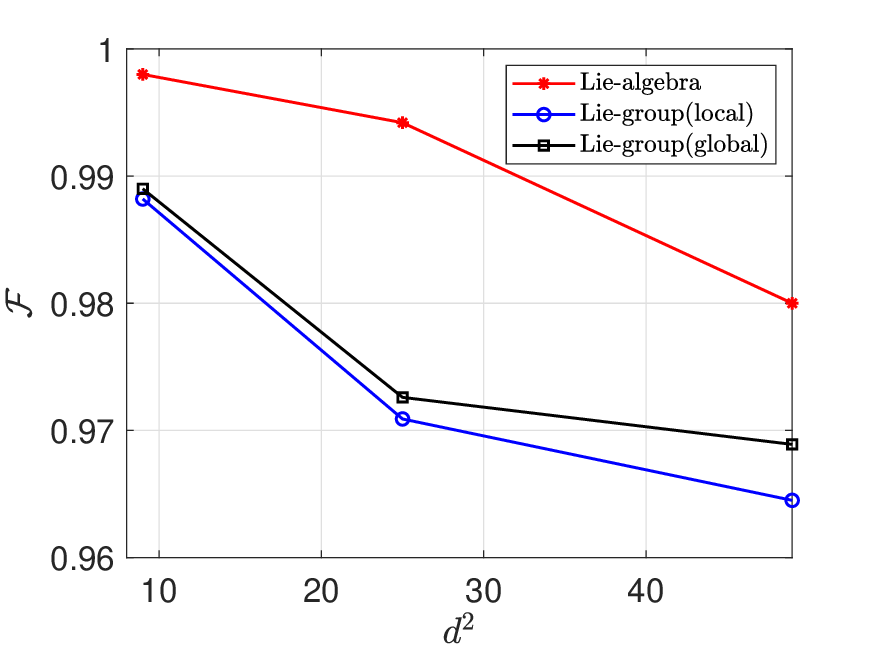}
	\caption{A comparison of the optimized fidelity, $\mathcal{F}$ of the CPhase gate achieved for the Lie algebraic and Lie group approaches (including both local single-qudit control and only global control)  is plotted as a function of the total Hilbert space dimension $d^2$, for the qudits of dimension $d=3,5,$ and $d=7$. For all of these simulations, we have taken the parameters given in Fig.~{\ref{fig:Fidelity figure 1}}.}
	\label{fig:comparison_approach}
\end{figure}

A comparison of the fidelities achieved based on the Lie algebraic and Lie group approaches is given in Fig.~\ref{fig:comparison_approach} for $d=3,5,$ and $d=7$. 
The results suggest that the Lie algebraic protocol slightly outperforms the Lie group protocol in the presence of decoherence.

\begin{figure}
\centering
	\includegraphics[width=0.5\textwidth]{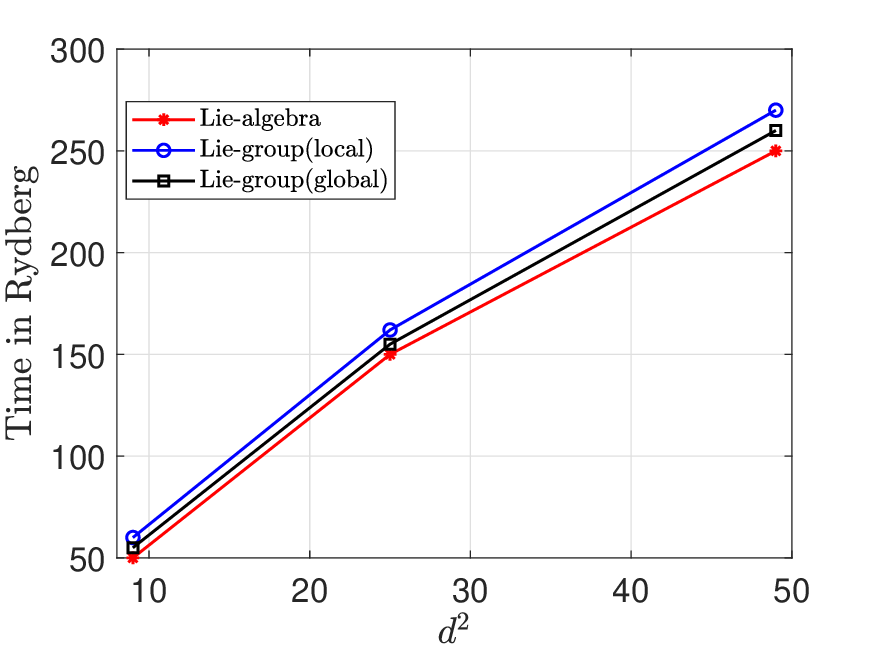}
	\caption{A comparison of the minimum time spent in the Rydberg state to implement the  CPhase gate achieved for the Lie algebraic and Lie group approaches (including both local single-qudit control and only global control)  is plotted as a function of the total Hilbert space dimension $d^2$, for the qudits of dimension $d=3,5,$ and $d=7$.
 For all of these simulations, we have taken the parameters given in Fig.~{\ref{fig:Fidelity figure 1}}.
 Thus the time required for the Lie algebraic control is smaller than the Lie group control which in turn contributes to the fidelity.}
	\label{fig:comparison_approach_time}
\end{figure}

{This difference in the performance can be attributed to the time spent in the Rydberg state for these two approaches, as shown in Fig.~\ref{fig:comparison_approach_time}. Fundamentally, we can understand this from the fact that the Lie algebraic approach has more control parameters as compared to the Lie group protocol. 
Thus, based on the Magnus expansion~\cite{merkel2009quantum,jurdjevic1972control,brockett1973lie}, the nested commutators which are at the heart of controllability become easier to achieve.  Both approaches yield high fidelities in large dimensional qudits.
Nevertheless, the Lie group approach may be preferable when considering the complexity necessary for experimental control}.

{ In general, a key experimental consideration for the successful implementation of open-loop quantum control is the effect of uncertainties in Hamiltonian parameters. These can be mitigated to some degree using the tools of robust quantum control \cite{anderson2015accurate, goerz2015optimizing, glaser2015training, koch2016controlling}.  Such techniques are generalizations of spin-echo type composite pulses which can be useful when there is sufficient coherence time.  With a detailed  understanding of the dominant inhomogeneities, robust optimal control can be used to implement suitable composite waveforms for  qudit entanglers on any platform.

The specific experimental foundation of this proposal is well-motivated by existing literature, particularly the work of the Jessen group~\cite{anderson2013unitary}.  
One particular issue discussed above is the trap-induced differential light shifts between the ground state and excited state $\TPT$ manifold \cite{PhysRevResearch.5.013219}.
It will be necessary to mitigate motional dephasing arising from vector- and tensor-shifts, which induce an $m_F$-dependence on polarizability, thus inducing possible motional dephasing between $m_F$ levels.
The easiest way around this problem is to operate with a linearly-polarized optical trap, with polarization vector aligned at the ``magic angle'' \cite{PhysRevX.8.041054} and corresponding magic wavelength \cite{doi:10.1126/science.1148259} for the $\SSZ \rightarrow \TPT$ transition.
This allows intra-state coherence within the $\TPT$ $F=9/2$ (and other $F$-levels) manifold, and inter-state (\textit{i.e.}, optical qubit) coherence between the $\SSZ$ and $\TPT$ $F=9/2$. We can also mitigate motional effects via high-fidelity ground-state cooling \cite{kaufman2012cooling, thompson2013coherence, lester2014raman}. }

\section{Conclusion and Outlook} 
\label{sec:conclusions_and_future_work}

Quantum computation with qudits has potential advantages when compared with architectures employing qubits. Implementing gates for qudit-based quantum computation is fundamentally more challenging, as the generators for these gates are not native Hamiltonians on physical platforms. One way to overcome this challenge is to use the tools of quantum optimal control, whereby we combine native Hamiltonians with time-dependent waveforms that drive the system in order to implement a universal gate set with high fidelity. 

In this work, we introduced two classes of numerical methods of quantum optimal control for implementing the qudit entangling gates, an essential component of the universal gate set. The first approach is based on continuous-time driving given a controllable Hamiltonian with tunable parameters and uses the Lie algebraic structure of the control problem. The second approach is more ``digital," using the Lie group structure to design a family of unitary maps that can be applied in sequence to achieve any nontrivial entangling gate of interest. 

As a specific example, we studied encoding a qudit in the nuclear spin of $^{87}$Sr, a species of atoms that is particularly important in quantum information processing. The nuclear spin can accommodate a qudit of dimension $d\le 10$. We have previously studied protocols for implementing single-qudit gates in SU($d$). To implement entangling gates we studied how we make two atoms interact using the well-known Rydberg blockade mechanism, and in particular, we studied Rydberg dressing schemes. Using this we are able to generate any two-qudit entangling gate, both using the Lie algebraic and Lie group based approaches.

We also studied how the fundamental effects of decoherence introduced by the finite lifetime of the Rydberg states reduce the gate fidelity. To model this we used a nonHermitian Hamiltonian and found that even when including decoherence, one could achieve high fidelity for these qudit entanglers. Given the flexibility of arbitrary control, we can seek the best  approach to encoding qudits and mitigating errors. 

Finally, while we have studied a particular case study in the context of neutral-atom quantum computing, the general methods we have developed here can be applied in other platforms,  including trap ions transmon qudits, and nanomagnets \cite{petiziol2021counteracting,chiesa2021embedded}, which also have natural encoding and control Hamiltonians.

\begin{acknowledgements}
This work was supported by the Laboratory Directed Research and Development program of Los Alamos National Laboratory under project numbers 20200015ER and 20210116DR,
and the NSF Quantum Leap Challenge Institutes program, Award No. 2016244.
The authors acknowledge fruitful discussions with Sri Datta Vikas Buchemmavari, Milad Marvian, Pablo Poggi, Jonathan Gross, Irfan Siddiqi, and Noah Goss   during various stages of this work. 
\end{acknowledgements}

\appendix

\clearpage
\section{Hyperfine structure of Rydberg states and Clebsch-Gordan coefficients}
\begin{figure}
\centering
\includegraphics[width=0.45\textwidth]{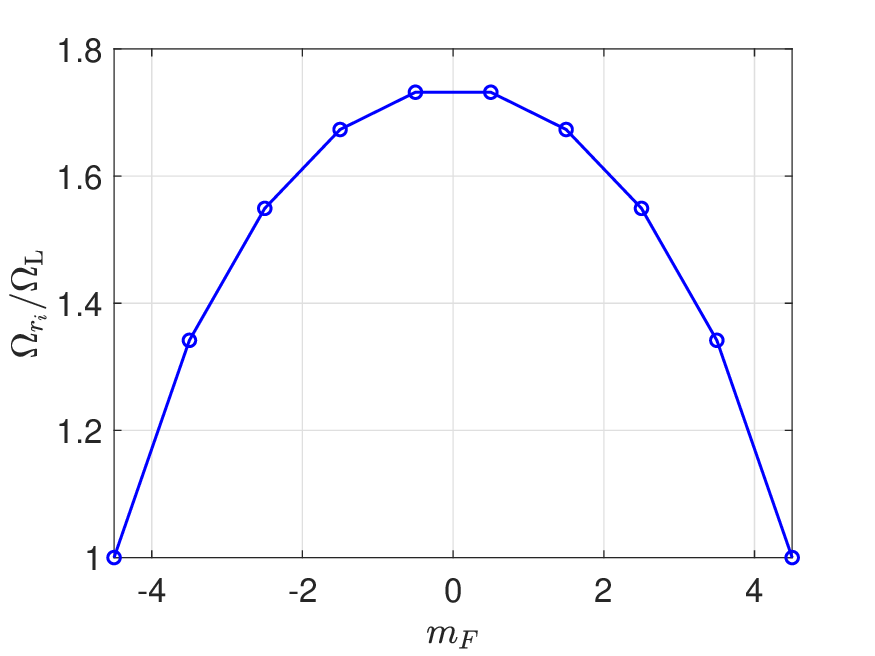}
\caption{Relative Rabi frequency, $\Omega_{r_i}/\Omega_{\mathrm{L}}$, plotted as a function of $m_F$ for $\pi$ polarized light for the $(5s5p)^3P_2F=9/2 \to (5sns) ^3S_1 F'=11/2$ transition to the Rydberg state.
The quadratic function arises due to the tensor polarizability. }
\label{fig:Omega_eff}
\end{figure}
As described in the Sec.~\ref{sec:controllability}c, to create entanglement we promote the population from the ground state $^{1}S_0$ to the first excited  $^3P_2$ state, with the hyperfine quantum number  $F=9/2$, and then consider a UV laser to excite the atoms to the $^{3}S_1$ Rydberg series to implement the interaction between atoms with adiabatic dressing (see Fig.~\ref{fig:set_up_figure}). The Rabi frequency characterizing the coupling of the different $m_F$ levels in the $^3P_2$ hyperfine manifold to the $^{3}S_1$ Rydberg states will be different due to the  Clebsch-Gordon Coefficients for these transitions. 
Let $\Omega_{\mathrm{L}}$ be the Rabi frequency on the $\ket{0_a}\to \ket{0_r}$ ($m_F=-9/2$ transition). 
The Rabi frequency experienced by the other levels is then
\begin{small}
\begin{equation}
      \Omega_{r_i}=\frac{\bra{F,m_F=-9/2+i}\ket{1,0; F', m_F=-9/2+i}}{\bra{F,m_F=-9/2}\ket{1,0; F', m_F=-9/2}}\Omega_{\mathrm{L}}, 
    \label{eq:Omega_eff}
\end{equation}
\end{small}
where we have chosen $F=9/2$ and $F'=11/2$, and  a $\pi$-polarized light. In Fig.~\ref{fig:Omega_eff} the Rabi frequencies of the different levels are given as a function of $m_F$, whose parabolic shape describes the tensor light shift, thus giving a natural nonlinearity which arises solely due to well-defined hyperfine structure of $^{87}$Sr.


\begin{figure}
\centering
		\includegraphics[width=0.49\textwidth]{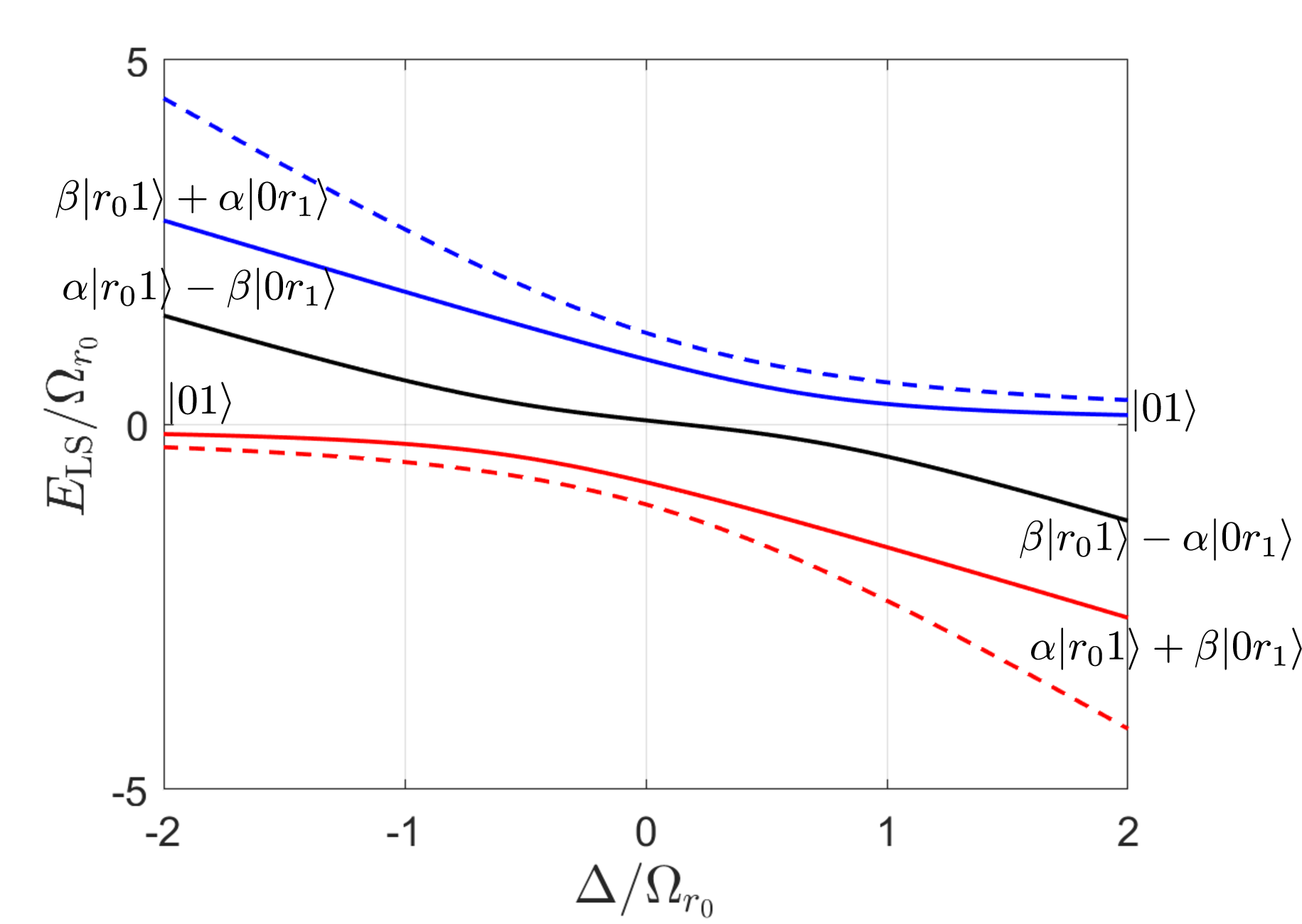}
	\caption{Autler-Townes splitting of the three dressed states as a function of detuning for the Hamiltonian in  Eq.  (\ref{eq:two_atom_Hamiltonian}), where $i=0, \;j=1$, such that $\ket{0} \equiv \ket{^3P_2, \; m_F=9/2}$ and $\ket{1} \equiv \ket{^3P_2, \; m_F=7/2}$. 
    Here $\alpha=\sqrt{7/16}$ and $\beta=\sqrt{9/16}$.
	The dashed line shows the AC Stark shift (light shift) in the absence of a perfect Rydberg blockade. The blue curve adiabatically connects to the clock states for large blue detuning and the red curve for large red detuning. The black curve is a dressed superposition that does not adiabatically connect to the clock states. The dashed lines show the light shifts in the absence of van der Waals interactions between the atoms. The difference between the solid line and the dashed line is the entangling power of the Hamiltonian $H_{2}^{12}$ defined in Eq.(\ref{eq:two_atom_Hamiltonian}).}
	
	\label{fig:eig_figure}
\end{figure}


Consider the Rydberg dressing scheme in Fig.~\ref{fig:set_up_figure}. 
In the perfect blockade regime, the two-atom Hamiltonian coupling of two magnetic sublevels labeled $i$ and $j$ is described by a three-level system, governed by the Hamiltonian,
\begin{equation}
\begin{aligned}
    H_2^{ij}=&-\Delta_i\ket{r_i j}\bra{r_i j}+\frac{\Omega_{r_i}}{2}\left(\ket{r_ij}\bra{ij}+\ket{ij}\bra{r_ij}\right)\\
    & -\Delta_j\ket{ir_j}\bra{ir_j}+\frac{\Omega_{r_j}}{2}\left(\ket{ir_j}\bra{ij}+\ket{ij}\bra{ir_j}\right),
        \label{eq:two_atom_Hamiltonian}
    \end{aligned}
\end{equation}
where $\Delta_i$ determines the detunings due to the differential Zeeman shit. Fig.~\ref{fig:eig_figure} shows the resulting AC Stark shifts on the three dressed states after diagonalizing this Hamiltonian. The dressed ground state is shown in red; the other two dressed states represent Autler-Townes splitting. In the absence of the van der Waals interaction the AC Stark shift (light shift) is the sum of the light shifts of each atom independently (dashed line in Fig.~\ref{fig:eig_figure}. The difference between these is the entangling energy. 

One can understand the entangling power of the Hamiltonian by studying the properties of the dressed energy levels as a function of detuning. Figure~\ref{fig:eig_figure} shows the particular case of $i=0$, $j=1$ for the Hamiltonian in Eq.~(\ref{eq:two_atom_Hamiltonian}), where  $\ket{0} \equiv \ket{m_F=9/2}$ and $\ket{1} \equiv \ket{m_F=7/2}$.
On the red side of detuning and for large detuning, as we start with the bare state and we adiabatically sweep through resonance, the state maps to the superposition of the two Rydberg states. 
Note, this is not an equal superposition as seen in \cite{mitra2020robust} due to the fact that the states $\ket{0}$ and $\ket{1}$ couple with different Rydberg Rabi frequency and detuning to the Rydberg states.

\section{Controllability}
\label{sec:controllability_of_the_Hamiltonian}
The quantum system is said to be controllable if, given a time-dependent Hamiltonian $H[\mathbf{c}(t)]$, there exist a time-dependent set of waveforms $\mathbf{c}(t)$, such that the one can generate an arbitrary unitary map. Here we consider those two-qudit unitary maps generated by an entangling Hamiltonian that is symmetric under the exchange of the qudits and thus does not require local addressing. To show that a Hamiltonian is controllable, we use the operator basis of irreducible spherical tensors on spin $j$  defined as \cite{sakurai2014modern,klimov2008generalized},
\begin{equation}
T^{(k)}_q=\sqrt{\frac{2k+1}{2j+1}}\sum_q \braket{j,k+q}{k,1;j,m}\ketbra{j,m+q}{j,m}.
    \label{eq:spherical_tensor}
\end{equation}
These satisfy the fundamental commutation relations,
\begin{equation}
    \begin{aligned}
    \comm{j_z}{T^{(k)}_q}=&q T^{(k)}_q,\\
    \comm{j_{\pm}}{T^{(k)}_q}=&\sqrt{k(k+1)-q(q\pm 1)}T^{(k)}_{q\pm 1}.
    \label{eq:spherical_tensor_commutators}
    \end{aligned}
\end{equation}
The set of operators $T^{(k)}_q$ form a complete orthonormal operator basis. Merkel \emph{et al.}~\cite{merkel2009quantum} showed that given a generating set of Hamiltonians $\{h_i\}$, if 
\begin{equation}
    \Tr{h_i,T^{k}_q}\neq 0
\end{equation}
for $k>2$, the system is fully controllable. That is, the set generates the whole Lie algebra of interest, which thus allows us to implement an arbitrary unitary map on the spin of the system using quantum control. 

\begin{figure}
\centering
\includegraphics[width=0.48\textwidth]{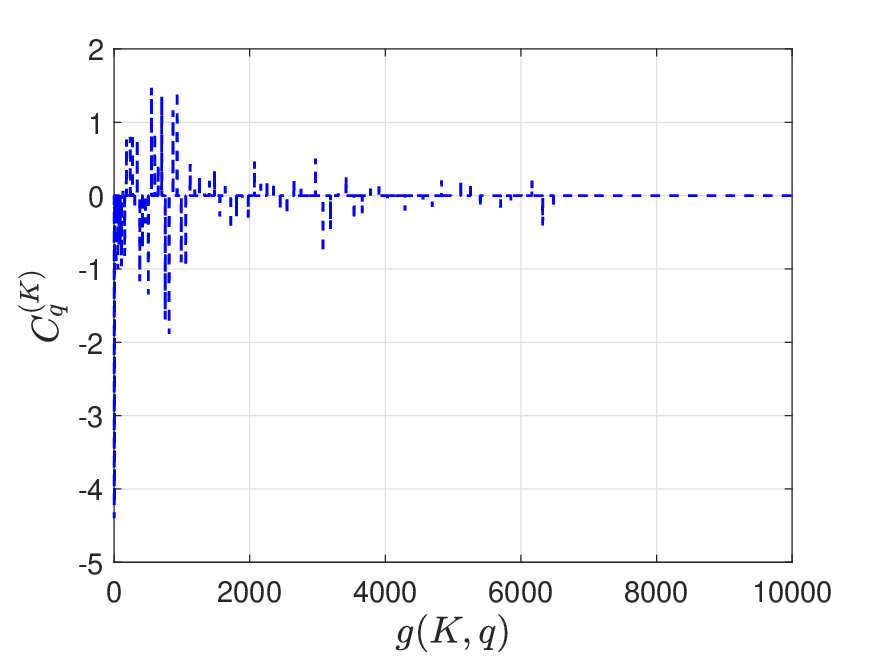}
\caption{The decomposition the entangling Hamiltonian $H_{\mathrm{ent}}$, Eq.~(\ref{eq:entangling_Hamiltonian}) in different orders of spherical tensors, $T^{(K)}_q$, for $j=99/2$,  an operator basis of dimension $D=2j+1=100$, spanning the two-qudit space for $d=10$. The expansion coefficients are given by $C^{(K)}_q= \left| {\rm Tr}(H_{\rm{ent}}T^{(K)\dag}_q)\right|^2$. We have ordered the expansion coefficients according to $g(K,q)=(k+1)^2-1+q$, where $0\leq k\leq j$, and $-k\leq q\leq k$. The existence of contributions from higher-rank tensors makes the system controllable when combined with time-dependent rf-fields that act locally on the atoms.}
\label{fig:controllability}
\end{figure}

We consider two-qudit systems, where the relevant Lie Group is $\mathrm{SU}(d^2)$; here $d^2=100$. We expand the entangling Hamiltonian in the operator basis of spherical tensors with $j=99/2$, spanning the space of dimension $D=2j+1=100$. Fig.~\ref{fig:controllability} shows operator decomposition of the entangling Hamiltonian $H_{\mathrm{ent}}$ in different orders of spherical tensors. One can see in this figure that there are contributions from higher rank tensors,  making the system controllable.

\section{Creating other symmetric qudit entanglers for the Lie algebraic approach}
\label{sec:Molmer_sorenson_gate}
\begin{figure*}

\includegraphics[width=0.98\textwidth]{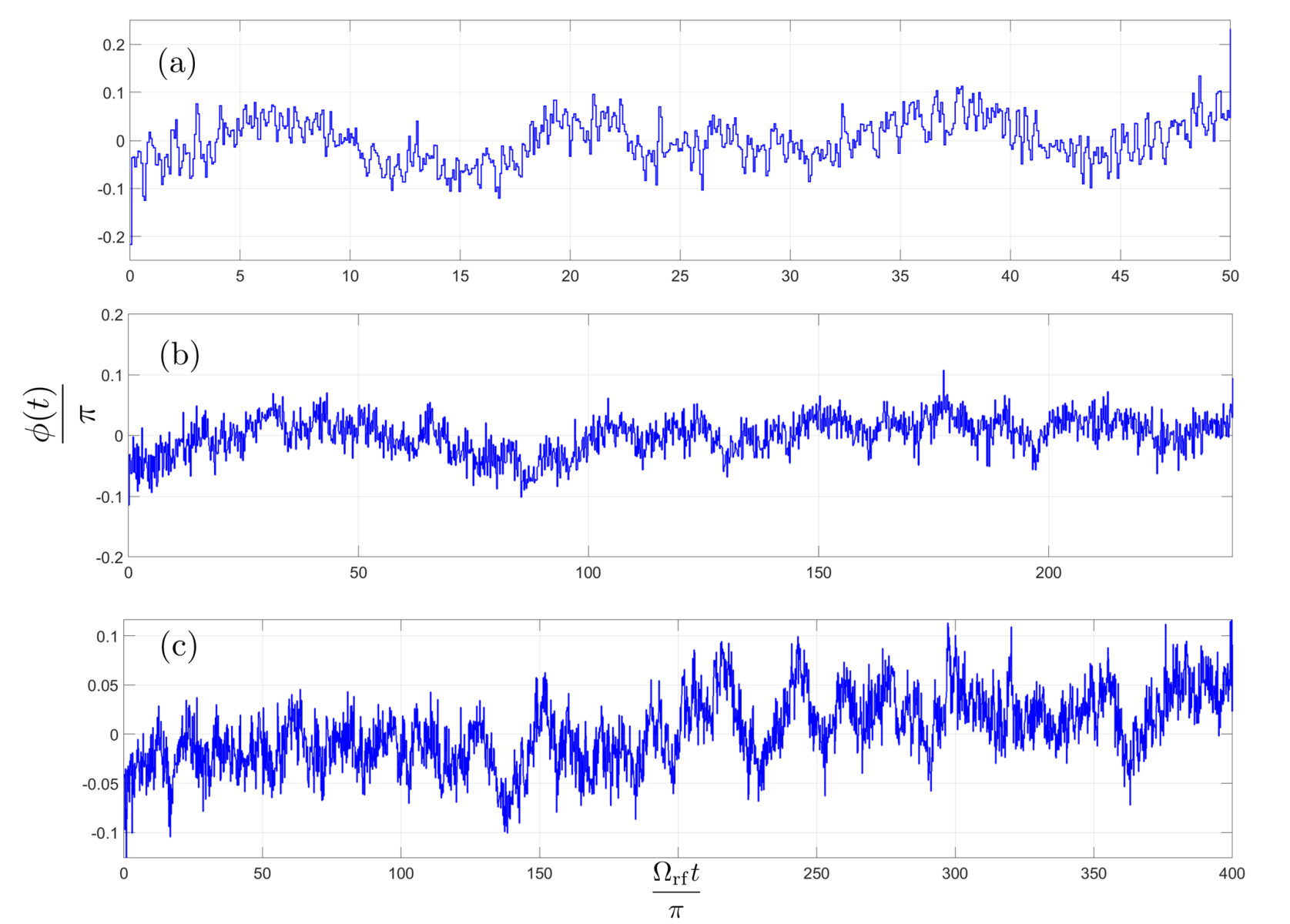}

    
    \caption{The figure gives the $\phi(t)$ that generates the M$\o$lmer-S$\o$renson gate as a function of time for $ \theta=\pi/2$ using the piecewise constant quantum control approach for the Hamiltonian given in Eq.\eqref{eq:entangling_Hamiltonian}. 
    In (a) the case of the $d=3$ for a total time of $\Omega_{\mathrm{rf}}T=50 \pi$ with $700$ piecewise constant steps. 
    In (b) the case of the $d=5$for a total time of $\Omega_{\mathrm{rf}}T=240 \pi$ with $1600$ piecewise constant steps.
    And in (c)  the case of the $d=7$for a total time of $\Omega_{\mathrm{rf}}T=240 \pi$ with $2500$ piecewise constant steps. 
    For all of these calculations we have taken $\Omega_{\mathrm{L}}=6 \Omega_{\mathrm{rf}} $.}
    \label{fig:Qudits_simulations_ms_gate}
\end{figure*}
Since the Hamiltonian described in Eq.\eqref{eq:entangling_qudit_Hamiltonian_1} can be used to create any symmetric two-qudit Hamiltonian, we can also generate the M$\o$lmer-S$\o$renson gate for qudits defined as, 
\begin{equation}
U_{\mathrm{MS}}(\theta)=\exp(-i\theta\frac{J_z^2}{2}).
\end{equation}
where the total angular momentum operator for the two qudits is
\begin{equation}
 J_z =\mathds{1}\otimes j_z + j_z \otimes \mathds{1}. 
\end{equation}

We employ the same procedure for optimal control as we discussed in the main text in designing the waveforms to implement the CPhase gate. Numerical examples of the waveforms that create the M$\o$lmer-S$\o$renson gate for $\theta=\pi/2$ are given in Fig.~\ref{fig:Qudits_simulations_ms_gate}. 
The figure shows $\phi(t)$, the piecewise constant of the control waveform, obtained using the GRAPE algorithm. Fig.~\ref{fig:Qudits_simulations}(a) shows the case of the $k=3$ the qutrit encoded in $d=10$. The total time is $T=50 \pi/\Omega_{\mathrm{rf}}$ and we divide the time into $700$ time steps for the quantum control. In Fig.~\ref{fig:Qudits_simulations}(b) we plot an example waveform for the case of the $d=5$ into our $10$ level system. We have a total time of $T=240 \pi/\Omega_{\mathrm{rf}}$ and we divide the time into $1600$ time steps for the quantum control. 
In Fig.~\ref{fig:Qudits_simulations}(c) we plot an example for the case of the $d=7$ into our $10$ level system. 
We have a total time of $T=400 \pi/\Omega_{\mathrm{rf}}$ and we divide the time into $2500$ time steps for the quantum control.

\bibliography{reference}
\end{document}